\documentclass[paper]{ieice}
\usepackage{graphicx}
\usepackage{amsmath} 
\usepackage{amsfonts} 
\usepackage[usenames, dvipsnames]{color}

\newtheorem{theorem}{Theorem}
\newtheorem{lemma}{Lemma}
\newtheorem{remark}{Remark}
\newcommand{\bbE}{\mathbb{E}}

\field{}
\title{Posterior Matching for Gaussian 
Broadcast Channels with Feedback}
\authorlist{
\authorentry[lantruong@u.nus.edu]{Lan\hspace{1mm}V. TRUONG}{n}{labelA}
\authorentry[hirosuke@ieee.org]{Hirosuke YAMAMOTO}{f}{labelB}
}
\affiliate[labelA]{The author is with National University of Singapore.}
\affiliate[labelB]{The author is with The University of Tokyo.}

\begin{document}
\maketitle
\begin{summary}
In this paper, the posterior matching scheme proposed by Shayevits and Feder is extended to the Gaussian broadcast channel with feedback, and the error probabilities and achievable rate region are derived for this coding strategy by using the iterated random function theory. A variant of the Ozarow-Leung code for the general two-user broadcast channel with feedback can be realized as a special case of our coding scheme. Furthermore, for the symmetric Gaussian broadcast channel with feedback, our coding scheme achieves the
linear-feedback sum-capacity like the LQG code and outperforms the Kramer code.
\end{summary}
\section{Introduction}\label{section1}
The capacity region of the broadcast channel with $M$ users (i.e.~$M$ receivers) is a well-known open problem. However, it is known that feedback can increase the capacity region for broadcast channels. Specially, Ozarow and Leung \cite{key1} proved for $M=2$ that feedback can increase the capacity region of the additive white Gaussian broadcast channel (AWGN-BC)  by cooperation between the users and the sender via feedback. Kramer \cite{key2} extended this coding scheme to the case of $M\geq 3$. Later, Elia \cite{key3} showed  for $M=2$ that the achievable rate region obtained by Ozarow and Leung \cite{key1} can be enlarged by using robust control theory. Ardestanizadeh et al.~\cite{key4} proposed a coding scheme based on LQG (Linear Quadratic Gaussian) control approach for the symmetric AWGN-BC with feedback, and showed that their LQG code can attain the same achievable rate region as the Elia scheme \cite{key3} for $M=2$ and outperforms the Kramer code \cite{key2} for the symmetric AWGN-BC with feedback for $M\geq 3$. The LQG code is derived based on a mapping from a feedback control problem to a linear code for the AWGN-BC with feedback. The achievable rate region is determined by the eigenvalues of the open-loop matrix of a linear system and the power constraint of channel input is related to the minimum power needed to stabilize the system using a feedback control signal.

Recently, Amor \emph{et al.}~\cite{key24},~\cite{key25} showed that the rate regions achieved by linear feedback coding schemes over dual multi-antenna AWGN multi-access channels (MACs) and broadcast channels (BCs) with independent noises coincide, and 
the sum-rate achieved by the LQG code is optimal among all the linear-feedback coding schemes for the symmetric AWGN-BCs. This optimal sum-rate is called linear-feedback sum-capacity, 
and they showed for $M=2$ that the linear-feedback sum-capacity of the scalar AWGN-BC with independent noises can be achieved by a simple rearrangement of Ozarow's MAC coding scheme~\cite{key-Oz}. (Refer to Remark~\ref{amor} 
in Section~\ref{section5} for more details.)
However, it is not shown for $M\geq 3$ how to construct a coding scheme for AWGN-BCs with feedback by a rearrangement of a coding scheme for AWGN-MACs with feedback. Note that since Kramer's MAC coding scheme~\cite{key2}, which is a generalization of Ozarow's MAC coding scheme for $M\geq 3$, uses  complex modulation coefficients, it is not easy to construct a BC coding scheme from Kramer's MAC coding scheme even if we try to use a rearrangement similar to the one used in \cite{key25}. 

In a more general setting, Gaspar et al.~\cite{key5}, \cite{key6} proposed a coding scheme for the AWGN-BC with correlated noises in the case of $M=2$ with arbitrary noise covariance and in the case of $M\geq 3$ such that the noise of each user is a multiple of the same Gaussian noise. For example, they showed that for all noise correlations other than $\pm 1$, the gap between the sum-rate of their scheme and the full-cooperation bound vanishes as the signal-to-noise ratio tends to infinity. Although their coding scheme works well in the asymptotic regime, it does not work well when the input power is not sufficiently large.

Shayevits and Feder \cite{key7} proposed the Posterior Matching (PM) Scheme for the point-to-point communication system with feedback, and they showed that the PM Scheme reduces to the Schalkwijk-Kailath scheme \cite{key8} when the channel is Gaussian. But, it is very hard to directly apply their scheme to AWGN-BCs because we need to assign multiple messages to a single vector and to refine the vector sequentially based on feedback to reduce the uncertainty of every user at the same time. To execute such a behavior, a higher order kernel is required for the reversed iterated function system (RIFS) used in the decoders, and all the decoders must know all the other decoders' messages. On the other hand, the indirect assignment methods used in the Ozarow-Leung code \cite{key1} or the Kramer code \cite{key2} can lead to only a suboptimal sum-rate compared with the Elia code \cite{key3} and the LQG code \cite{key4} as mentioned above. Specially, they assumed that the transmitted signal at each time $n$ is a linear combination of different signal components, each of which is intended to decrease the uncertainty of each user, and they also imposed a redundant restriction such that each signal component at time $n+1$ must be statistically independent of the signal feedbacked from the corresponding user at time $n$. This idea is originated from the Schalkwijk-Kailath scheme \cite{key8} and repeated in the Shayevitz-Feder scheme \cite[Section A]{key7} to attain the capacity for point-to-point AWGN-BCs with feedback. But for the AWGN-BCs with feedback, this scheme cannot realize so good performance as the Elia code \cite{key3} and the LQG code \cite{key4}.

In this paper, we extend the PM scheme \cite{key7} to AWGN-BCs with $M$ users 
by devising a new encoding scheme for any $M$ such that an $M\times M$ binary Hadamard matrix
exists. Our encoding procedure can be considered as an optimization of the Kramer scheme \cite{key2} by using some mathematical tricks. The decoding scheme uses the same technique 
as the Shayevits-Feder scheme \cite{key7}. But our coding scheme is a general one for AWGN-BCs with feedback because it includes all the coding schemes treated in \cite{key1} and \cite{key2} as special cases, and we derive the achievable rate region of the proposed coding scheme. Then, we prove that a variant of the Ozarow-Leung scheme \cite{key1} obtained from our scheme can achieve the same achievable rate region as the original Ozarow-Leung scheme. Furthermore, we propose a coding scheme for physically non-degraded symmetric AWGN-BCs with feedback which can achieve the linear-feedback sum-capacity like the LQG code. Besides, since our coding scheme is a variant of the Kramer code, it has a potential to achieve not only the asymptotic capacity \cite{key5}, \cite{key6} but also a good performance in non-asymptotic settings. More precisely, we can determine the code length (i.e.~the repetition number of feedback) necessary to attain a given target of error probabilities and coding rates in our coding scheme in the same way as other PM schemes. This is an advantage over the Elia code \cite{key3} and the LQG code \cite{key4}, in which we cannot determine the necessary code length because the decoding error exponent and achievable mean square error exponent are treated only in the asymptotic setting for these codes. 

This paper is organized as follows. Section \ref{section2} presents the channel model and some mathematical preliminaries. A general time-varying coding scheme is proposed for AWGN-BCs with feedback in Section \ref{section3}, and the achievable rate region and error probabilities for this general scheme are derived in Section \ref{section4}. Section \ref{section5} shows that a variant of the Ozarow-Leung coding scheme can be obtained from our coding scheme. We show that the proposed coding scheme can achieve the linear-feedback sum-capacity for physically non-degraded symmetric AWGN-BCs with feedback in Section \ref{section6}. 
Finally in Section \ref{section8}, we compare the sum-rate for the AWGN-BC with the one for the AWGN-MAC. 
\section {Channel Model and Preliminaries} \label {section2}
\subsection{Mathematical Notations} \label{subsection21}
Upper-case letters and lower-case letters denote random variables and their realizations, respectively. A real-valued random variable $X$ is associated with a distribution $\mathbb{P}_X(\cdot)$ defined on the usual Borel $\sigma$-algebra over $\mathbb{R}$, and we write $X \sim \mathbb{P}_{X}$. The cumulative distribution function (c.d.f.)~of $X$ is given by $F_X(x)=\mathbb{P}_X((-\infty,x])$, and their inverse c.d.f is defined as $F_X^{-1}(t)\equiv \mbox{inf}\{x:F_X(x) > t\}$. The uniform probability distribution over $(0,1)$ is denoted by $\mathcal{U}$. In addition, we use the following notation. $(f\circ g)(x)\equiv f(g(x))$, ${\bf Y}_p^{q(m)}\equiv (Y_p^{(m)}, Y_{p+1}^{(m)},...,Y_q^{(m)})$ for $p\leq q$, and $\mbox{tr}({\bf A})$ is the trace of matrix ${\bf A}$. In this paper, we use the following lemma:
\begin{lemma}[{{\cite[Lemma 1]{key7}}}] \label{bc:lemma1}
Let $X$ be a continuous random variable with $X\sim \mathbb{P}_X$ and $\Theta$ be a uniform distribution random variable, i.e. $\Theta \sim \mathcal{U}$, and $X$ be statistical independent of $\Theta$. Then $F_X^{-1}(\Theta) \sim \mathbb{P}_X$ and $F_X(X) \sim \mathcal{U}$. 
\end{lemma} 


The binary Hadamard matrix \cite{key9} of order $M$ is an $(M\times M)$ matrix of $+1$s and $-1$s such that ${\bf H}_M{\bf H}_M^T=M{\bf I}$ where ${\bf I}$ is the $(M\times M)$ identity matrix. 
It is not yet known for which values of $M$ an ${\bf H}_M$ exists. 
However, we know that if the Hadamard matrix of order $M$ exists then $M$ is $1, 2, 4$, or a multiple of $4$. Moreover, if $M$ is of the form $2^m$ for a positive integer $m$ we can construct ${\bf H}_M$ by using Sylvester's method. In addition,  Paley's construction, which uses quadratic residues, can be used to construct Hadamard matrices of order $M$ when $M$ is 
equal to $p+1$ for a prime $p$ and $M$ is also a multiple of $4$. 

\subsection{AWGN-BCs with Feedback} \label{subsection22}
\begin{figure}[t!]
	\centering
		\includegraphics[width=0.48\textwidth]{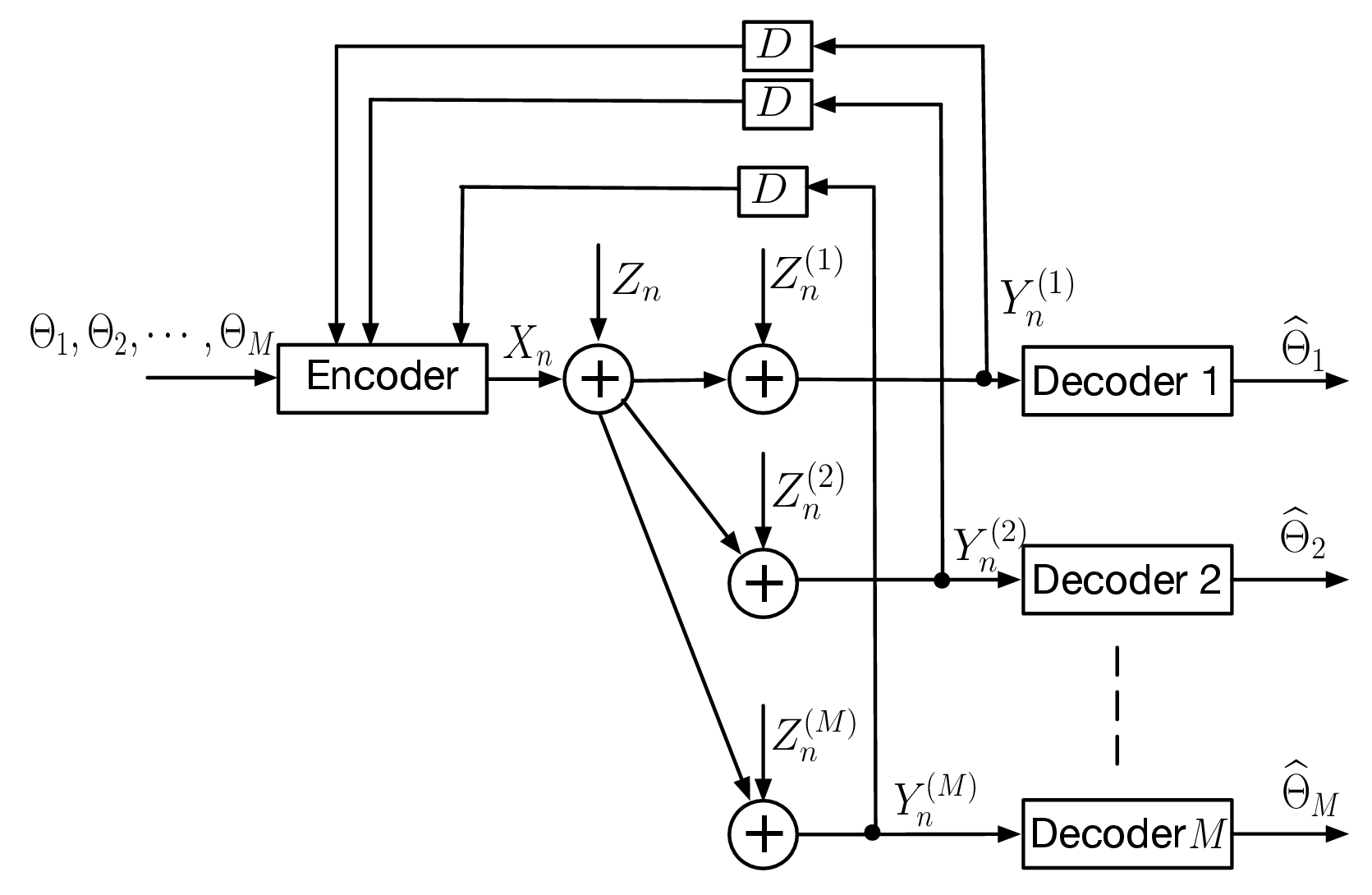}
	\caption{$M$-user Gaussian Broadcast Channel with Feedback}
	\label{fig:Broadcastchannel}
\end{figure}

We extend the communication model treated in \cite{key1} to the case of AWGN-BCs. Consider the communication system shown in Fig.~\ref{fig:Broadcastchannel} such that one encoder and $M$ decoders are connected via an AWGN-BC and all channel outputs are noiselessly feedbacked to the encoder. Let $\Theta_m$ be a random message point uniformly distributed over the unit interval that must be transmitted from the encoder to decoder $m \in \{1,2,...,M\}$.
At each time $n$, the received signal of decoder $m$ is 
\begin{align}
\label{eq1}
Y^{(m)}_n=X_n + Z_n + Z_n^{(m)},
\end{align}
where $X_n \in \mathbb{R}$ is the symbol transmitted from the encoder at time $n$, and $Y^{(m)}_n \in \mathbb{R}$ is the signal received by decoder $m$ at time $n$. $Z_n$ is a common white Gaussian noise  with variance $\sigma^2$, and $Z_n^{(m)}$ are individual white Gaussian noises with variance $\sigma_m^2$ for $m=1, 2, \cdots, M$.  
For physically non-degraded AWGN-BCs, we can set $\sigma^2=0$ and $\sigma_m^2>0$. We also assume that output symbols are casually feedbacked to the encoder and the transmitted symbol $X_n$ at time $n$ can depend on both messages $(\Theta_1, \Theta_2, ..., \Theta_M)$ and the previous channel output sequences $\left({\bf Y}^{n-1(1)},{\bf Y}^{n-1(2)},\cdots,{\bf Y}^{n-1(M)}\right)$ where ${\bf Y}^{n-1(m)}\equiv(Y^{(m)}_1,Y^{(m)}_2,\cdots,Y^{(m)}_{n-1})$. 

An \emph{encoding scheme} for an AWGN-BC is a measurable transmission function $g_n: (0,1)^M \times \mathbb{R}^{(n-1)M}\rightarrow \mathbb{R}$, so that the channel input 
generated by the encoder is given by
\begin{align}
\label{eq2}
X_n=g_n\left(\Theta_1, ..., \Theta_M, {\bf Y}^{n-1(1)},{\bf Y}^{n-1(2)},...,{\bf Y}^{n-1(M)}\right).
\end{align}

A \emph{decoding rule} for the AWGN-BC is the sequences of measurable mappings $\{\Delta_n^{(m)}: \mathbb{R}^n \rightarrow \mathcal{E}\}_{n=1}^{\infty}$, where $\mathcal{E}$ is the set of all open intervals in $(0,1)$. We refer to $\Delta_n^{(m)}({\bf y}^{n(m)})$ as the 
decoded interval of decoder $m$.  The error probabilities at time $n$ are defined as
\begin{align}
\label{eq3}
p_{n,e}^{(m)}\equiv\mathbb{P}(\Theta_m \notin \Delta_n^{(m)}({\bf Y}^{n(m)}))
\end{align}
for $m =1, 2, \cdots, M$, and the corresponding coding rate at time $n$ is defined by
\begin{equation}
\label{eq4}
R_n^{(m)}\equiv-\frac{1}{n}\log\left|\Delta_n^{(m)}\left({\bf Y}^{n(m)}\right)\right|,
\end{equation} where $\left|\Delta^{(m)}_n({\bf Y}^n)\right|$ is the length of the interval $\Delta^{(m)}_n({\bf Y}^n)$.

We say that a coding scheme achieves a rate tuple $(R_1, R_2,\cdots, R_M)$ 
over an AWGN-BC if for all $m\in \{1,2,...,M\}$, it satisfies
\begin{align}
\lim_{n\rightarrow \infty}\mathbb{P}\left(R^{(m)}_n<R_m\right)&=0, \label{eq5-1} \\
\label{eq5}
\lim_{n \rightarrow \infty}p_{n,e}^{(m)}&=0.
\end{align}
The rate tuple is achieved within an input power constraint $P$ if it also satisfies
\begin{equation}
\label{eq6}
\limsup_{n\rightarrow \infty}\frac{1}{n}\sum_{k=1}^n \bbE[X_k^2] \leq P.
\end{equation}

An \emph{optimal fixed rate} decoding rule for an AWGN-BC with feedback for rate tuple $\left(R_1,R_2,...,R_M\right)$ is the one that decodes the tuple of fixed length intervals $(J_1,J_2,...,J_M)$ satisfying $|J_m|=2^{-nR_m}$ for each $m$, which maximizes each marginal posteriori probability, i.e.,
\begin{align}
\label{eq7}
\triangle^{(m)}_n({\bf y}^{n(m)})=\underset{{J_m \in \mathcal{E}: |J_m|=2^{-nR_m}}}{\mbox{argmax}} \mathbb{P}_{\Theta_m|{\bf Y}^{n(m)}}(J_m|{\bf y}^{n(m)}).
\end{align}

An \emph{optimal variable rate} decoding rule with target error probabilities $p_{n,e}^{(m)}=\delta^{(m)}_n$ is the one that decodes the tuple of minimal-length intervals $(J_1,J_2,...,J_M)$ such that each accumulated marginal posteriori probability exceeds corresponding target, i.e.,
\begin{align}
\label{eq8}
\triangle^{(m)}_n({\bf y}^{n(m)})=\underset{{J_m\in \mathcal{E}: \mathbb{P}_{\Theta_m|{\bf Y}^{n(m)}}(J_m|{\bf y}^{n(m)})\geq 1- \delta^{(m)}_n}}{\mbox{argmin}}|J_m|.
\end{align}
Both decoding rules make good use of the marginal posterior distribution of the message point $\mathbb{P}_{\Theta_m|Y^n}$ which can be calculated online at the encoder and each decoder.  Refer \cite{key7} for more details.  Then, the following lemma holds.
\begin{lemma}[{{\cite[Lemma 3]{key7}}}] \label{lem-2} The achievability defined by 
\eqref{eq5-1}--\eqref{eq6} implies the achievability in the standard framework.
\end{lemma}

\begin{remark}
In the standard framework, a message $i_m$ uniformly distributed over $\{1,2,...,2^{n\tilde{R}_n^{(m)}}\}$ is sent to decoder $m$ via a BC  when the coding rate is $\tilde{R}_n^{(m)}$.
It is shown in the proof of \cite[Lemma 3]{key7} that
if $\tilde{R}_n^{(m)}$ satisfies 
\begin{align}
\tilde{R}_n^{(m)} \leq R_m +\frac{1}{n} \log \left(1-\sqrt{p_{e,n}^{(m)}}-\tau_n\right)
\end{align}
for some $\tau_n>0$ such that $\lim_{n\rightarrow \infty}\tau_n=0$, 
then we can choose message points $\theta_{i_m,n}$ in $(0,1)$ such that $\theta_{i_m+1,n}-\theta_{i_m,n}\geq 2^{-nR_m}$ for  $1\leq i_m \leq 2^{n\tilde{R}_n^{(m)}}-1$ 
and the decoding error probability $\tilde{p}_{e,n}^{(m)}$ in the standard framework is upper bounded by
\begin{align}
\tilde{p}_{e,n}^{(m)}< \sqrt{p_{e,n}^{(m)}}.
\end{align}
 Note that the encoding in the standard framework can be realized by mapping each message $i_m$ to $\theta_{i_m,n}$. 
 Hence, if $R_m$ is achievable in the sense of this section,
then $R_m$ is also achievable in the meaning of  the standard framework.
See \cite{key4} and \cite{key7} for the details of the proof of Lemma \ref{lem-2}.
Also note that since $M$ independent message points $(\Theta_1, \Theta_2, \cdots, \Theta_M)$ are used in the encoding function $g_n$ defined by \eqref{eq2}, 
each message $i_m$ can be mapped to the message point $\theta_{i_m,n}$ independently from other messages $i_{m'}$, $m'\neq m$. Therefore, Lemma \ref{lem-2} holds for the case of BCs in the same way as the case of point-to-point communication treated in \cite{key7}.
\end{remark}

\section{A Time-varying Coding Scheme for AWGN-BCs with Feedback}\label{section3}

In this section, we propose \emph{a time-varying coding scheme} for AWGN-BCs with feedback. 
\subsection{Encoding Scheme}\label{subsection31}
Assume that the sender wants to send $M$ messages $\{\Theta_m\}_{m=1}^M$ to $M$ users, respectively, where $\Theta_m$ satisfying $\Theta_m \sim \mathcal{U}$ is the message for user $m$ and $\Theta_m$ is independent of  $\Theta_{m'}$ for $m'\neq m$.

\vspace{0.3cm}
\noindent{\bf Initialization at $n=1$.}\footnote{We use $M$ time slots for the initialization. But for simplicity of notation, $n=1$ is assigned for these $M$ time slots.}   

For each $m, 1\leq m \leq M$: 
\begin{itemize}
\item The encoder broadcasts a message $S_1^{(m)} = F_{S}^{-1}(\Theta_m)$, where $S \sim \mathcal{N}(0,P_0)$, and $P_0>0$ is determined based on the channel situation.
\item User $m$ receives $Y_1^{(m)} = S_1^{(m)} + Z_1+ Z_1^{(m)}$ and feedbacks $Y_1^{(m)}$ to the encoder.
 \end{itemize}

\noindent{\bf Recursion for $n \geq 2$.}
\begin{itemize}
\item The encoder creates a random variables $S_n^{(m)}$ defined by
\begin{equation}
\label{eq9}
S_n^{(m)} = \frac{1}{a_{n-1}^{(m)}} \left(S_{n-1}^{(m)} - b_{n-1}^{(m)} Y_{n-1}^{(m)}\right),
\end{equation}
where $a_{n-1}^{(m)}>0$ and $b_{n-1}^{(m)}, m=1,2,...,M,$ are real numbers which are also chosen based on the channel situation.
\item The encoder broadcasts the following signal to all the users:
\begin{equation}
\label{eq10}
X_{n}=  \beta_n \sum_{m=1}^M \alpha_{n}^{(m)} S_n^{(m)}.
\end{equation}
Here, $\beta_n$ is a real number, which is chosen to satisfy the input power constraint~\eqref{eq6}, and
\begin{align}
\label{eq11}
{\boldsymbol \alpha}_{n}=[\begin{array}{cccc}\alpha_{n}^{(1)}&\alpha_{n}^{(2)}&\cdots&\alpha_{n}^{(M)}\end{array}]^T
\end{align}
is a modulated vector.\\
\item User $m$ receives the signal
\begin{equation}
\label{eq12}
Y_n^{(m)}=  \beta_n \sum_{m=1}^M \alpha_n^{(m)} S_n^{(m)} + Z_n+ Z_n^{(m)}, 
\end{equation}
and it feedbacks $Y_n^{(m)}$ to the encoder.
\end{itemize}

\subsection{Decoding Scheme}\label{subsection32}
\noindent{\bf Recursion for  $n\geq 2$:}
\begin{itemize}
\item Each user $m$ receives $Y_n^{(m)}$ given by~\eqref{eq12}.
\item Each user $m$ selects a fixed interval $J_1^{(m)}=(s_m, t_m) \subset \mathbb{R}$ with respect to $S_n^{(m)}$. 
\item Then, each user $m$ estimates the interval $J_n^{(m)}$ for the $S_1^{(m)}$ as follows.
\begin{align}
\label{eq13}
J_n^{(m)}=\left(T_n^{(m)}(s_m), T_n^{(m)}(t_m)\right)
\end{align}
where
\begin{align}
\label{eq14}
T_n^{(m)}(x)\equiv w_1^{(m)} \circ w_2^{(m)} \cdots \circ w_n^{(m)}(x)
\end{align}
and
\begin{equation}
\label{eq15}
w_n^{(m)}(x)\equiv a_n^{(m)} x + b_n^{(m)} Y_n^{(m)}.
\end{equation}
Note that $a_n^{(m)}>0$ ensures that $w_n^{(m)}(x)$ and $T_n^{(m)}(x)$ are monotonically increasing in $x$ for any realization ${\bf y}^{n(m)}$ of ${\bf Y}^{n(m)}$.
 \item Finally, the decoded interval $ \Delta_n^{(m)}({\bf Y}^{n(m)})$ is determined for $\Theta_m$ as follows:
 \begin{equation}
\label{eq16}
 \Delta_n^{(m)}({\bf Y}^{n(m)})\equiv F_{S}\left(J_n^{(m)}\right),
\end{equation}
 where $S \sim \mathcal{N}(0,P_0)$, and for the p.d.f.~$f_S(t)$ of $S$,
 \begin{equation}
\label{eq17}
 F_S ((a, b)) \equiv \left(\int_{-\infty}^a f_{S}(x) dx,  \hspace{2mm} \int_{-\infty}^b f_S(x) dx \right).
 \end{equation}
 \end{itemize} 

 \section{Error Analysis for the Time-varying Coding Scheme for AWGN-BCs with Feedback}\label{section4}
In this section, we evaluate the performance of the time-varying posterior matching scheme proposed in Section \ref{section3}. \\
\begin{theorem}
\label{bc:thm1}
The time-varying coding scheme for the AWGN-BC given by Fig.~\ref{fig:Broadcastchannel} achieves any rate tuple $(R_1, R_2,...,R_M)$ if it satisfies
\begin{align}
\label{eq18}
R_m < R_m^*\equiv - \limsup_{n\rightarrow \infty} \log a_n^{(m)}
\end{align}
for $0< \limsup_{n\rightarrow \infty} a_n^{(m)}  < 1$ and $W^{(m)}_n\equiv\bbE[S_n^{(m)}]^2$ is upper bounded. Furthermore, the error probability $p_{n,e}^{(m)}$ satisfies that for every $m \in \{1,2,...,M\}$\footnote{$f_1(n)=o(f_2(n))$ means that $\lim_{n\rightarrow\infty}f_1(n)/f_2(n)=0$.},
\begin{align}
\label{eq19}
-\log p_{n,e}^{(m)}=o\left(2^{2n(R_m^*-R_m)}\right).
\end{align}
\end{theorem}
\begin{remark} 
Eq.~\eqref{eq19} means that $p_{n,e}^{(m)}$ can go to zero in the following way\footnote{$\exp_2(n)\equiv 2^{n}$.}:
\begin{align}
p_{n,e}^{(m)}&\leq \kappa \exp_2\left(-\frac{2^{2n(R_m^*-R_m)}}{u(n)}\right)\nonumber\\
& =\kappa  \exp_2\left(-2^{2n(R_m^*-R_m-\frac{1}{2n}\log u(n))}\right)
\end{align}
for some $\kappa>0$ and any $u(n)$ satisfying that $\lim_{n\rightarrow\infty}u(n)=\infty$.
Hence,  if we use $u(n)$ satisfying   $\lim_{n\rightarrow\infty} (1/2n)\log u(n) <\eta (R_m^*-R_m)$ 
for some $\eta$, $0<\eta<1$, then $p_{n,e}^{(m)}$ can go to zero with double exponential order.
More precislely, $\kappa$ can be determined from \eqref{eq25}.
\end{remark}
\emph{Proof} Let $R_n^{(m)}$ be the instant rate to transmit message $\Theta_m$ to 
 user $m$. For any fixed rate $R_m$, we have
\begin{align}
\mathbb{P}\left(R_n^{(m)}< R_m \right) &\stackrel{(a)}{=} \mathbb{P}\left(-\frac{1}{n} \log\left|\Delta_n^{(m)}({\bf Y}^{n(m)})\right|  < R_m \right) \nonumber  \\
&=\mathbb{P}\left(|\Delta_n^{(m)}({\bf Y}^{n(m)})| > 2^{-nR_m}\right)  \nonumber \\
\label{eq20}
&\stackrel{(b)}{\leq} \mathbb{P}\left( |J_n^{(m)}|> 2^{-n R_m}/ K\right)
\end{align}
where 
\begin{equation}
\label{eq21}
K =\sup_{x \in \mathbb{R}} \{f_S(x)\}.
\end{equation}
Here, (a) follows from \eqref{eq4},  and (b) holds from \eqref{eq16}, \eqref{eq17}, and \eqref{eq21}. 

Note from \eqref{eq15} that for all $t, s \in \mathbb{R}$,  we have
\begin{equation}
\label{eq22}
|w_n^{(m)}(t)- w_n^{(m)}(s)| = a_n^{(m)} |t-s|.
\end{equation}
For $a_m\equiv\limsup_{n\rightarrow \infty} a_n^{(m)}$ we have $R_m^* \equiv  \log a_m^{-1} > 0$ since $0<a_m<1$. Hence, for any rate $R_m < R_m^*$, we can find an $\epsilon >0$ such that $R_m < \log (a_m +\epsilon)^{-1}$ and $a_m +\epsilon <1$. Furthermore, there exists an $N_{\epsilon} \in \mathbb{N}$ such that $\sup_{n> N_{\epsilon}} a_n^{(m)} < a_m + \epsilon $.  Define $v_m\equiv\sup_{1\leq n \leq N_{\epsilon}} a_n^{(m)}$.  Then, from \eqref{eq20} and \eqref{eq22}, we have 
\small
\begin{align}
\mathbb{P}&\left(R_n^{(m)}< R_m \right) \nonumber\\
&\leq \mathbb{P}\left( |J_n^{(m)}|> 2^{-nR_m}/ K\right) \nonumber \\
&\stackrel{(a)}{\leq} K 2^{nR_m} \bbE\left[\bbE\left(|w_1^{(m)}\circ w_2^{(m)} \cdots \circ w_n^{(m)}(t_m) 
\right. \right. \nonumber \\
&\quad \left. \left. -w_1^{(m)}\circ w_2^{(m)} \cdots \circ w_n^{(m)}(s_m)|\hspace{1mm}\big|{\bf Y}_2^{n(m)}\right)\right] \nonumber  \\
&\stackrel{(b)}{\leq} K 2^{nR_m} v_m \bbE\left[|w_2^{(m)}\circ w_3^{(m)} \cdots \circ w_n^{(m)}(t_m)
\right.  \nonumber \\
&\quad \left. -w_2^{(m)}\circ w_3^{(m)} \cdots \circ w_n^{(m)}(s_m)|\right] \nonumber \\
&\hspace{1mm}\vdots \nonumber\\
&\stackrel{(c)}{\leq} K 2^{nR_m}v_m^{N_{\epsilon}} \bbE\left[|w_{N_{\epsilon}+1}^{(m)}\circ w_{N_{\epsilon}+2}^{(m)} \cdots \circ w_n^{(m)}(t_m)
\right.  \nonumber \\
&\quad \left. -w_{N_{\epsilon}+1}^{(m)}\circ w_{N_{\epsilon}+2}^{(m)} \cdots \circ w_n^{(m)}(s_m)|\right]  \nonumber  \\
&\hspace{1mm}\vdots \nonumber \\
\label{eq23}
&\stackrel{(d)}{\leq} K 2^{nR_m} v_m^{N_{\epsilon}} (a_m+\epsilon)^{(n-N_{\epsilon})}|J_1^{(m)}|,
\end{align}
\normalsize
where (a) follows from Markov's inequality and the law of iterated expectations, (b) follows from~\eqref{eq22} and $v_m\equiv\sup_{1\leq n \leq N_{\epsilon}} a_n^{(m)}$, (c) is the recursive application of (b), and (d) follows from $\sup_{n> N_{\epsilon}} a_n^{(m)} < a_m +\epsilon$ and the recursive applications of (b).

From \eqref{eq23} and $a_m +\epsilon <1$, it is easy to see that $\mathbb{P}(R_n^{(m)} < R_m) \rightarrow 0$ holds if
\begin{equation}
\label{eq24}
 |J_1^{(m)}| = o\left(2^{n(\log(a_m+\epsilon)^{-1} - R_m)}\right).
\end{equation}

For $Q(x)\equiv\int_x^{\infty}\frac{1}{\sqrt{2\pi}}e^{-t^2/2}dt$ and $W_n^{(m)}=\bbE[S_n^{(m)}]^2$, we obtain\footnote{$f_1(n)\sim f_2(n)$ means that $\lim_{n\rightarrow\infty} f_1(n)/f_2(n) =1$.}
\begin{align}
p_{n,e}^{(m)} &=\mathbb{P}\left(\Theta_m \notin \Delta_n^{(m)}\left({\bf Y}^{n(m)}\right)\right) \nonumber \\
&=\mathbb{P}\left(\Theta_m \notin F_S(J_n^{(m)})\right) \nonumber  \\
&\stackrel{(a)}{=}\mathbb{P}\left(S_1^{(m)} \notin J_n^{(m)}\right) \nonumber  \\
&=\mathbb{P}\left(S_n^{(m)} \notin J_1^{(m)}\right) \nonumber  \\
&\stackrel{(b)}{=} 2Q\left(\frac{|J_1^{(m)}|}{2\sqrt{W_n^{(m)}}}\right) \label{eq25-0} \\
&\stackrel{(c)}{\sim}\frac{2\sqrt{W_n^{(m)}}}{\sqrt{2\pi}|J_1^{(m)}|}\exp\left(-\frac{|J_1^{(m)}|^2}{8W_n^{(m)}}\right).\label{eq25}
\end{align}
Here, (a) follows from the fact that $\Theta_m$ is uniformly distributed over $(0,1)$ 
and this equality holds for any realization ${\bf y}^{n(m)}$ of the random vector ${\bf Y}^{n(m)}$.
(b) follows from the fact that  
$S_n^{(m)}$ is Gaussian with $\bbE[S_n^{(m)}]=0$, which can be shown inductively from 
\eqref{eq9} and \eqref{eq12},  and $J_1^{(m)}$ is symmetric if we set $s_m= -t_m$. 
(c) follows from that
$Q(x)$ satisfies
\begin{align}
\frac{1}{\sqrt{2\pi}x}\left(1-\frac{1}{x^2}\right) \exp\left(-\frac{x^2}{2}\right) < Q(x)\nonumber\\
<\frac{1}{\sqrt{2\pi}x} \exp\left(-\frac{x^2}{2}\right) \label{eq-Q}
\end{align}
 for any $x>0$.

From $R_m <\log(a_m+\epsilon)^{-1} < R_m^*$, we can select $J_1^{(m)}$ satisfying~\eqref{eq24} and $|J_1^{(m)}| \to \infty$ as $n\to \infty$. Furthermore, since $W_n^{(m)} $ is upper bounded by some $W$, we have
\begin{align}
\label{eq26}
\frac{|J^{(m)}_1|^2 }{8W^{(m)}_n} \geq \frac{|J^{(m)}_1|^2 }{8W} \rightarrow \infty.
\end{align}
More precisely by substituting~\eqref{eq24} into~\eqref{eq25}, 
$p_{n,e}^{(m)}$ satisfies 
\begin{align}
\label{eq27}
-\log p_{n,e}^{(m)} 
&\sim \frac{|J_1^{(m)}|^2}{8W_n^{(m)}}\log {\rm e}-\log \frac{2\sqrt{W_n^{(m)}}}{\sqrt{2\pi}|J_1^{(m)}|} \nonumber\\
&\sim  \frac{|J_1^{(m)}|^2}{8W_n^{(m)}}\log {\rm e}\nonumber\\
&= o\left(2^{2n(\log(a_m+\epsilon)^{-1} -R_m)}\right).
\end{align}
Since the above argument holds for any sufficiently small $\epsilon>0$, we can attain 
\begin{align}
-\log p_{n,e}^{(m)} = o\left(2^{2n(R_m^* -R_m)}\right).
\end{align}
\QED 
\begin{remark} Since we can estimate $R_m^*$ and know our desired rate $R_m$ in advance, it is possible to choose $\epsilon$ appropriately as a target. This means that the decoding algorithm is technically realizable.  However, there is a tradeoff between the transmission rate $R_m$ (the possible values of $\epsilon$) and the code length $n$. If $R_m$ is very close to $R_m^*$, $\epsilon$ must be very small. As a result, the required $N_{\epsilon}$ becomes very large. Furthermore, since $R_m$ is also very close to $\log (a_m +\epsilon)^{-1}$, the error probabilities $p_{n,e}^{(m)}$ decay slowly to zero. In the sequel, a very large code length $n$ is required if $R_m$ is close to $R_m^*$. On the contrary, if $R_m^*-R_m$ is large, we can choose quite large $\epsilon$, which makes the required $N_{\epsilon}$ smaller and the decay of error probabilities faster. 
\end{remark}
\begin{remark} \label{rmk-re}
In the case of finite $n$,  $\mathbb{P}(R_n^{(m)}< R_m)$ is not zero even if 
$J_m$ satisfies \eqref{eq24}. But this does not worsen the error probability $p_{n,e}^{(m)}$ if  retransmission is allowed.
Note that since  the encoder obtains ${\bf y}^{n(m)}$ via the feedback channel, both the encoder and decoder $m$ can know the value of $R_n^{(m)}$ for  ${\bf y}^{n(m)}$. 
Hence, 
they can know whether event $\{R_n^{(m)}< R_m\}$ occurred or not when they received ${\bf y}^{n(m)}$. If  event $\{R_n^{(m)}< R_m\}$ occurs, they discard this transmission and resend the same massage $i_m$ in the standard framework.  
This retransmission decreases the coding rate  of massage $i_m$  from $\tilde{R}_n^{(m)}$ to 
$\tilde{R}_n^{(m)} (1 - \mathbb{P}(R_n^{(m)}< R_m))$. But, this degradation of coding rate is negligible if $\mathbb{P}(R_n^{(m)}< R_m)$ is sufficiently small. 
\end{remark}
\begin{remark}
If we cannot use the retransmission described in Remark \ref{rmk-re}, event $\{R_n^{(m)}< R_m\}$ makes a decoding error. In this case, we need to minimize the total decoding error probability given by 
$p_{n,e}^{(m)}+ \mathbb{P}(R_n^{(m)}< R_m)$, and hence we cannot attain double exponential order.
By setting $|J_1^{(m)}|^2(\log {\rm e})/8W= n(\log(a_m+\epsilon)^{-1}-R_m)$
 in \eqref{eq23} and \eqref{eq25},  the error exponent of the total error probability is given by
\begin{align}
\lim_{\epsilon\rightarrow0}\lim_{n\rightarrow\infty}&\left[-\frac{1}{n} \log \left(p_{n,e}^{(m)}+ \mathbb{P}(R_n^{(m)}< R_m)\right)\right] \nonumber\\
 &\geq\lim_{\epsilon\rightarrow0}\left[\log(a_m+\epsilon)^{-1}-R_m\right]\nonumber\\
 &= R_m^*-R_m.\label{eq-new32}
\end{align}
\end{remark}
\section{A Variant of the Ozarow-Leung Coding Scheme for Two-User AWGN-BCs with Feedback}\label{section5}
Denote 
\begin{equation}
\label{eq28}
\rho_n\equiv\frac{ \bbE[S_n^{(1)} S_n^{(2)}]}{P/2}.
\end{equation}
In this case, we set
\begin{align}
P_0&= P/2,\\
\alpha_n^{(1)}&= 1,\\
\alpha_n^{(2)} &= g\hspace{1mm} \mbox{sgn}(\rho_n).
\end{align} 
Here, $\mbox{sgn}(x)\equiv1$ if $ x\geq 0$ and $\mbox{sgn}(x)\equiv-1$ if $x<0$. $g$ is a nonnegative number which allows a trade-off between $R_1^*$ and $R_2^*$ \cite{key1}. We also define
\begin{align}
\label{eq32}
\beta_n &= \sqrt{\frac{2}{1+g^2 + 2g|\rho_n|}},\\
\label{eq33}
a_n^{(1)}&= \sqrt{\frac{\mbox{var}(S_n^{(1)}|Y_n^{(1)})}{P/2}}, \\
\label{eq34}
a_n^{(2)}&= \sqrt{\frac{\mbox{var}(S_n^{(2)}|Y_n^{(2)})}{P/2}}, \\
\label{eq35}
b_n^{(1)}&= \frac{\bbE[S_n^{(1)}Y_n^{(1)}]}{\mbox{var}(Y_n^{(1)})}, \\
\label{eq36}
b_n^{(2)}&= \frac{\bbE[S_n^{(2)}Y_n^{(2)}]}{\mbox{var}(Y_n^{(2)})}.
\end{align}
By substituting \eqref{eq33}--\eqref{eq36} into \eqref{eq9}, we can show for $m=1$ and 2 that
\begin{align}
\label{eq37}
S_{n+1}^{(m)} &= F_{S}^{-1}\circ F_{S_n^{(m)}|Y_n^{(m)}}(S_n^{(m)}|Y_n^{(m)}).
\end{align} (see \cite{key10}, \cite{key11}).

From Lemma~\ref{bc:lemma1}, each realization $y_n^{(m)}$ of $Y_n^{(m)}$ satisfies that
\begin{align}
\label{eq38}
F_{S_n^{(m)}|Y_n^{(m)}}(S_n^{(m)}|y_n^{(m)}) \sim \mathcal{U},
\end{align}
which means
\begin{align}
\label{eq39}
F_{S_n^{(m)}|Y_n^{(m)}}(S_n^{(m)}|Y_n^{(m)}) \sim \mathcal{U}.
\end{align}
Since $S\sim \mathcal{N}(0,P_0)=\mathcal{N}(0,P/2)$, we have $S_1^{(m)}=F_S^{-1}(\Theta_m)\sim \mathcal{N}(0,P/2)$ from Lemma~\ref{bc:lemma1}. Repeating this procedure we obtain
\begin{equation}
\label{eq40}
S_n^{(m)} \sim \mathcal{N}(0,P/2)
\end{equation} for any $n\geq 1$. In addition, we have from \eqref{eq10} and \eqref{eq12} that
\begin{align}
X_n &= \beta_n[S_n^{(1)}\alpha_n^{(1)} + S_n^{(2)}\alpha_n^{(2)} ] \nonumber \\
\label{eq49new}
&= \beta_n[S_n^{(1)} + g\hspace{1mm}\mbox{sgn}(\rho_n)S_n^{(2)} ],\\
Y_n^{(1)}&=\beta_n[S_n^{(1)} +g \hspace{1mm}\mbox{sgn}(\rho_n)S_n^{(2)} ] + Z_n+  Z_n^{(1)},\\
Y_n^{(2)}&=\beta_n[S_n^{(1)} +g\hspace{1mm} \mbox{sgn}(\rho_n)S_n^{(2)} ]+ Z_n + Z_n^{(2)}.
\end{align}
Since $S_n^{(m)}$ satisfies $\bbE[S_n^{(m)}]=0$ from \eqref{eq40}, we have
\begin{align}
\bbE[X_n]&=\bbE[Y_n^{(1)}]= \bbE[Y_n^{(2)}]=0.
\end{align}
Furthermore, from \eqref{eq40} we also have $\bbE[(S_n^{(m)})^2]=P/2$. Hence,
\begin{align}
\label{eq45}
\bbE[X_n^2]&= P,  \\
\label{eq46}
\bbE[S_n^{(1)}Y_n^{(1)}]&=(P/2) \beta_n(1+g|\rho_n|),\\ 
\label{eq47}
\bbE[S_n^{(2)}Y_n^{(2)}]&=(P/2)\beta_n\mbox{sgn}(\rho_n)(g+|\rho_n|),\\
\label{eq48}
\mbox{var}(Y_n^{(1)}) &=P+\sigma^2 + \sigma_1^2,\\
\label{eq49}
\mbox{var}(Y_n^{(2)})&= P+\sigma^2 + \sigma_2^2.
\end{align}
Note that the following relations hold. (Refer, e.g. \cite[page 323]{key16}.)
\begin{align}
\label{eq50}
\mbox{var}(S_n^{(1)}|Y_n^{(1)})&= \mbox{var}(S_n^{(1)}) - \frac{(\bbE[S_n^{(1)}Y_n^{(1)}])^2}{\mbox{var}(Y_n^{(1)})},\\
\label{eq51}
\mbox{var}(S_n^{(2)}|Y_n^{(2)})&= \mbox{var}(S_n^{(2)}) - \frac{(\bbE[S_n^{(2)}Y_n^{(2)}])^2}{\mbox{var}(Y_n^{(2)})}.
\end{align}

Substituting \eqref{eq45}--\eqref{eq51} into \eqref{eq33}--\eqref{eq36}, we finally have
\begin{align}
\label{eq52}
a_n^{(1)} &= \sqrt{\frac{\sigma^2 + \sigma_1^2 +(Pg^2(1-\rho_n^2))/(1+g^2+2g|\rho_n|)}{P+\sigma^2+\sigma_1^2}},  \\
\label{eq53}
a_n^{(2)} &= \sqrt{\frac{\sigma^2+ \sigma_2^2 + (P(1-\rho_n^2))/(1+g^2+ 2g|\rho_n|)}{P+\sigma^2 +\sigma_2^2}}, 
\\
\label{eq54}
b_n^{(1)} &= \frac{(P/2)\beta_n(1+g|\rho_n|)}{P+\sigma^2+\sigma_1^2},\\
\label{eq55}
b_n^{(2)} &= \frac{(P/2)\beta_n\mbox{sgn}(\rho_n)(g+|\rho_n|)}{P+\sigma^2+\sigma_2^2}.
\end{align}

From \eqref{eq9} for $m=1$ and 2, we have 
\begin{align}
\label{eq56}
\bbE[S_{n+1}^{(1)}S_{n+1}^{(2)}]&=\frac{1}{a_n^{(1)}a_n^{(2)}}\left(\bbE[S_n^{(1)}S_n^{(2)}] - b_n^{(1)} \bbE[S_n^{(2)}Y_n^{(1)}] \right. \nonumber \\
&\quad \left. - b_n^{(2)} \bbE[S_n^{(1)}Y_n^{(2)}] + b_n^{(1)}b_n^{(2)} \bbE[Y_n^{(1)}Y_n^{(2)}]\right).
\end{align}
By substituting \eqref{eq28} and \eqref{eq52}--\eqref{eq55} into \eqref{eq56} and some calculations, $\rho_n$ must satisfy
\begin{align}
\label{eq57}
&\rho_{n+1}\nonumber\\
&=\frac{A \rho_n -\frac{PB}{D(|\rho_n|)}(g+|\rho_n|)(1+g|\rho_n|)\mbox{sgn}(\rho_n)}{\sqrt{A}\sqrt{\left(\sigma^2+\sigma_1^2+\frac{Pg^2(1-\rho_n)^2}{D(|\rho_n|)}\right)\left(\sigma^2+\sigma_2^2+\frac{P(1-\rho_n^2)}{D(|\rho_n|)}\right)}},
\end{align}
where
\begin{align}
A&=(P+\sigma^2+\sigma_1^2)(P+\sigma^2+\sigma_2^2), \\
B&=P+\sigma^2+\sigma_1^2+\sigma_2^2,  \\
D(x)&=1+g^2 +2gx.
\end{align}

It is very difficult to affirm that the sequence $|\rho_n|$ is convergent. One strategy to overcome this difficulty is to keep $|\rho_n|$ unchanged (see \cite{key1}). 
Hence,  we set $\rho_n=(-1)^{n+1}\rho$, where $\rho$ is the biggest solution in $(0,1)$ of the following equation :
\begin{align}
\label{eq61}
x+&\frac{A x -\frac{PB}{D(x)}(g+x)(1+gx)}{\sqrt{A}\sqrt{\left(\sigma^2+\sigma_1^2+\frac{Pg^2(1-x)^2}{D(x)}\right)\left(\sigma^2+\sigma_2^2+\frac{P(1-x^2)}{D(x)}\right)}}\nonumber\\
&=0.
\end{align}
Note that \eqref{eq61} has a solution in $(0,1)$ since the left hand side of \eqref{eq61} is negative at $x=0$ and positive at $x=1$. 

Then, we have from \eqref{eq52} and \eqref{eq53} that
\begin{align}
&\limsup_{n\rightarrow \infty} a_n^{(1)} \nonumber\\
&= \sqrt{\frac{\sigma^2 + \sigma_1^2 +(Pg^2(1-\rho^2))/(1+g^2+2g\rho)}{P+\sigma^2+\sigma_1^2}},\\
&\limsup_{n\rightarrow \infty} a_n^{(2)} \nonumber\\
&= \sqrt{\frac{\sigma^2 + \sigma_2^2 +(P(1-\rho^2))/(1+g^2+2g\rho)}{P+\sigma^2+\sigma_2^2}}.
\end{align}
It is easy to verify that $0< \limsup_{n\rightarrow \infty} a_n^{(m)} < 1$ for $m=1$ and 2. Hence, from Theorem \ref{bc:thm1} the proposed scheme achieves any rate-pair ($R_1, R_2$) if
\begin{align}
R_1< R_1^* &= -\limsup_{n\rightarrow\infty} \log a_n^{(1)} \nonumber  \\ 
&= \frac{1}{2} \log\left( \frac{P+\sigma^2 + \sigma_1^2}{\sigma^2 +\sigma_1^2 +( Pg^2(1-\rho^2))/D(\rho)}\right),\\
R_2< R_2^* &= -\limsup_{n\rightarrow\infty} \log a_n^{(2)} \nonumber  \\
&= \frac{1}{2}\log \left( \frac{P+\sigma^2 + \sigma_1^2}{\sigma^2 +\sigma_2^2 +( P(1-\rho^2))/D(\rho)}\right).
\end{align} 
The error probabilities decay to zero as
\begin{align}
-\log p_{n,e}^{(1)} &= o\left(2^{2n(R_1^*-R_1)}\right),\\
-\log p_{n,e}^{(2)}&= o\left(2^{2n(R_2^*-R_2)}\right).
\end{align}
\begin{remark} The encoding scheme for $M=2$ treated in this section is a variant of the Ozarow-Leung coding scheme \cite{key1} which is represented by a form of time-varying posterior matching \cite{key10}, \cite{key11}. However, the performance of this code is worse than the one of the LQG code \cite{key4} and the Elia code \cite{key3}. Using the same approach,  we can obtain a variant of the Kramer code \cite{key2} for $M>2$. In the next section, we show that by choosing sequences $\alpha_n^{(m)}, b_n^{(m)}$ appropriately, we can achieve larger coding rate for $M\geq 2$. Specifically, we show that our proposed coding scheme for the symmetric AWGN-BCs with feedback attains 
the linear-feedback sum-capacity like the LQG code \cite{key4}, which is larger than the achievable sum-rate of the Kramer code \cite{key2}. 
\end{remark}
\begin{remark}
\label{amor}
The Amor-Steinberg-Wigger (ASW) coding scheme~\cite{key25}  for $2$-user asymmetric AWGN-BCs is constructed by a rearrangement of  the Ozarow coding scheme for $2$-user AWGN-MACs~\cite{key-Oz}, where two messages are assigned to two vectors with different powers and the power of each message can vary at each time $n$. See \cite[(189)]{key25}. But, the variant of the Ozarow-Leung coding scheme treated in this section uses a constant power at every time $n$ as shown in \eqref{eq40}.
Therefore, for the 2-user asymmetric case, our coding scheme is generally inferior to the ASW coding scheme. However, in the 2-user {\em symmetric} case, our coding scheme can attain the linear-feedback sum-capacity, like the ASW coding scheme, as shown in Section \ref{section6}.
We conjecture that by choosing appropriately $a_n^{(m)},b_n^{(m)},\beta_n$ in general setting given by \eqref{eq9},~\eqref{eq10}, our coding scheme can also attain the same coding rates as the ASW coding scheme for the 2-user asymmetric case. 
Furthermore, it is expected that our coding scheme can be extended to the $M$-user  asymmetric AWGN-BC channels with feedback easier than the ASW coding scheme because our scheme works for  {\em real} AWGN-BC channels, but Kramer's MAC coding scheme \cite{key2}, which is a generalization of the Ozarow MAC coding scheme, uses a {\em complex} modulation. These extensions are interesting future works.
\end{remark}

\section{$M$-user  Physically Non-degraded Symmetric AWGN-BC with Feedback}\label{section6}
In this section, we consider a physically non-degraded symmetric AWGN-BC with $\sigma^2_1=\sigma^2_2=...=\sigma^2_M=1$ and $\sigma^2=0$.
For this case, the following theorem holds.

\begin{theorem}
\label{bc:thm2}
For the $M$-user physically non-degraded symmetric AWGN-BC with feedback
satisfying $\sigma^2_1=\sigma^2_2=\cdots=\sigma^2_{M}=1$ and $\sigma^2=0$, 
the time-varying coding scheme proposed in Section~\ref{section3} can achieve 
the linear-feedback sum-capacity, i.e.~the sum-rate $R_{{\rm sum}}$ satisfying 
\begin{align}
R_{{\rm sum}}=\sum_{m=1}^M R^*_m = \frac{1}{2}\log\left(1+P\lambda \right), \label{eq-n-76}
\end{align}
where $\lambda$ is the biggest solution in $[1,M]$ of the following equation:
\begin{align}
\label{eq87}
(P\lambda+1)^{M-1}=[(P/M)\lambda(M-\lambda)+1]^M.
\end{align}
\end{theorem}
\vspace{0.3cm}

Theorem \ref{bc:thm2} will be proved in Section \ref{section7}.
The sum-rate given by \eqref{eq-n-76} coincides with the sum-rate of the LGQ code \cite[Theorem 2]{key4}, which is the linear-feedback sum-capacity of the symmetric AWGN-BC treated in this section \cite[Corollary 5]{key25}.

From this theorem, like the MAC case, we can prove that for large $M$,
\begin{align}
\sum_{m=1}^M R_m^* \approx \frac{1}{M}\log M +\frac{1}{2}\log\log M.
\end{align} 
Refer~\cite[(72)]{key2} for details. This means that the difference of the sum-rate of AWGN-BC with between feedback and no feedback grows as $(\log\log M)/2$ similar to the case of MACs.

\vspace{0.3cm}
Next we derive the tight upper bounds of  $p_{n,e}^{(m)}$ and $\mathbb{P}\left(R_n^{(m)}< R_m \right)$ for this symmetric case. 
Since $a_n^{(m)}$ can be fixed as $a_n^{(m)}=a$ for all $m$ and $n$ in this case as we will show in Section \ref{section7}, 
it holds in \eqref{eq22} that 
\begin{align}
|w_n^{(m)}(t)-w_n^{(m)}(s)|=a|t-s|.
\end{align} 
This means that we do not need to use $\epsilon$  in \eqref{eq23} in this case. Therefore, from \eqref{eq23} and \eqref{eq25-0}, if we choose 
\begin{align}
|J_1^{(m)}|=\frac{2\sqrt{W}2^{n(-\log a-R_m)}}{u(n)}=o\left(2^{n(-\log a-R_m)}\right)
\end{align} 
for any $u(n)$ and some constant $W$ such that $\lim_{n\rightarrow \infty}u(n)= \infty$ and $W_n^{(m)}\leq W$ for all $n$ and $m$, we can construct the coding scheme satisfying 
\begin{align}
p_{n,e}^{(m)}
&\leq 2Q\left(\frac{2^{n(-\log a-R_m)}}{u(n)}\right), \label{eq-pne}\\
\mathbb{P}\left(R_n^{(m)}< R_m \right)
&\leq K 2^{nR_m}  a^n |J_1^{(m)}| \nonumber\\
& =   \frac{2K \sqrt{W}}{u(n)}.
\end{align}

\begin{remark} 
In the symmetric case treated in this section, 
it holds from \eqref{eq-Q} and \eqref{eq-pne}
that  for $R^*\equiv-\log a$, 
\begin{align}
-\log p_{n,e}^{(m)}=o\left(2^{2n(R^*-R_m)}\right).
\end{align}
Furthermore, it also holds from \eqref{eq-new32} that
\begin{align}
\lim_{n\rightarrow\infty}\left[-\frac{1}{n} \log \left(p_{n,e}^{(m)}+ \mathbb{P}(R_n^{(m)}< R_m)\right)\right] \geq R^*-R_m.
\end{align}
\end{remark}

Since \eqref{eq-pne} gives the tight upper bound of $p_{e,n}^{(m)}$, 
we can know how many $n$ is required to achieve the targets
of $p_{n,e}^{(m)}$.
On the other hand, 
the LQG code satisfies  
\begin{align}
p_{n,e}^{(m)}\leq 4\times 2^{-2n(-\log a-R_m-\epsilon_n)}
\end{align} 
where $\epsilon_n \to 0$ as $n\to \infty$ \cite[(48)] {key4}.
However, we cannot know necessary $n$ for the targets of $p_{n,e}^{(m)}$ in this code
because the error exponent and achievable mean square error (MSE) exponents are only given in asymptotic settings.   The same holds for the Elia code \cite{key3}.

It is also worth noting that since our encoding scheme is a variant of the Kramer code, it has potential to achieve not only the symmetric capacity but also a good performance in asymmetric settings~\cite{key6}. But it is very difficult for the LQG approach to treat the asymmetric setting.

\section{Proof of Theorem~\ref{bc:thm2}.}\label{section7}
The following Lemmas~\ref{bc:lemma3} and~\ref{bc:lemma4} play important roles to prove 
Theorem \ref{bc:thm2}.
\begin{lemma}\label{bc:lemma3}
Let $\lambda^{(1)}, \lambda^{(2)}, ...,\lambda^{(M)}$ be a set of positive numbers satisfying: \\
\begin{align}
\label{eq68}
\lambda^{(m+1)} =\frac{1+(P/M)\lambda(M-\lambda)}{1+P\lambda}\lambda^{(m)}
\end{align}
for $m=1, 2, ..., M-1$, where $\lambda^{(1)}=\lambda$ is the biggest positive root of \eqref{eq87}.
Assuming that $\gamma$ is a negative number satisfying
\begin{equation}
\label{eq70}
\gamma \geq -\frac{\lambda}{P\lambda+1},
\end{equation}
then, we have $\lambda^{(m)} + \gamma > 0$ for all $m$.
\end{lemma}

\emph{Proof} From \eqref{eq87} and \eqref{eq68}, we have 
\begin{align}
\lambda^{(M)}&=\frac{[1+(P/M)\lambda(M-\lambda)]^{M-1}}{(1+P\lambda)^{M-1}}\lambda^{(1)} \nonumber \\
&=\frac{1}{1+(P/M)\lambda(M-\lambda)}\lambda^{(1)} \nonumber \\
\label{eq71}
&=\frac{\lambda}{1+(P/M)\lambda(M-\lambda)}
\end{align}
Combining \eqref{eq71} with \eqref{eq70}, we obtain 
\begin{align}
\lambda^{(M)} + \gamma &\geq \lambda^{(M)}  -\frac{\lambda}{P\lambda+1} \nonumber \\
&=\frac{\lambda}{1+(P/M)\lambda(M-\lambda)} -\frac{\lambda}{P\lambda+1} \nonumber \\
&=\frac{(P/M)\lambda^3}{(1+P\lambda)(1+(P/M)\lambda(M-\lambda))} > 0.
\end{align}
Moreover, from~\eqref{eq68} we have for all $m=1,2,...,M-1$ that
\begin{align}
\lambda^{(m+1)}&=\frac{1+P\lambda-(P/M)\lambda^2}{1+P\lambda}\lambda^{(m)}\nonumber \\
&\leq \lambda^{(m)}.
\end{align}
This means that
\begin{align}
\lambda^{(m)}+\gamma \geq \lambda^{(M)}+\gamma >0, \hspace{2mm} \forall m=1,2,...,M.
\end{align}
\QED
\begin{lemma}\label{bc:lemma4}
 For any positive number $\lambda$, the following simultaneous equations have a unique solution pair $(b, \gamma)$ in $b>0$.
\begin{align}
\label{eq75}
\gamma&=\frac{P\lambda +1 }{1+(P/M) \lambda (M-\lambda)}\left[\gamma+ \frac{M}{P} b^2\right],\\
\label{eq76}
\gamma &= \frac{1}{4b^2}\left[Mb^2 + \frac{(P/M)\lambda^2}{1+P\lambda}\right]^2 -\lambda.
\end{align}
Moreover, we have
\begin{align}
\label{eq77}
&0 > \gamma \geq -\frac{\lambda}{1+P\lambda},\\
\label{eq78}
Mb^2 &- 2b\sqrt{\lambda+\gamma} +\frac{(P/M)\lambda^2}{1+P\lambda} =0.
\end{align}
\end{lemma}

\emph{Proof} Eq.~\eqref{eq76} is equivalent to \eqref{eq78}, and \eqref{eq75} is equivalent to
\begin{equation}
\label{eq79}
\gamma =-\frac{(M/P)^2 b^2 (P\lambda+1)}{\lambda^2}.
\end{equation}

Substituting \eqref{eq79} into \eqref{eq76}, we have
\begin{align}
-\frac{(M/P)^2 b^2 (P\lambda+1)}{\lambda^2} = \frac{1}{4b^2}\left[Mb^2 + \frac{(P/M)\lambda^2}{1+P\lambda}\right]^2 -\lambda,
\end{align}
which means
\begin{align}
&\left[M^2 + 4 \frac{(M/P)^2 (P\lambda+1)}{\lambda^2}\right]b^4 - 2\left[\frac{P\lambda^2 + 2 \lambda}{1+P\lambda}\right] b^2 \nonumber \\ 
&\quad + \frac{(P/M)^2 \lambda^4}{(1+P\lambda)^2}=0.
\end{align}
Since the discriminant of the above quadratic equation is equal to zero, this equation has a unique solution $b^2$ given by 
\begin{align}
b^2 = \frac{(P\lambda^2 + 2\lambda)/(1+P\lambda)}{M^2 + 4 [(M/P)^2 (P\lambda+1)]/\lambda^2}.
\end{align}
Since we choose $b>0$ as the statement in Lemma \ref{bc:lemma4}, we get
\begin{align}
b=\sqrt{\frac{(P\lambda^2 + 2\lambda)/(1+P\lambda)}{M^2 + 4 [(M/P)^2 (P\lambda+1)]/\lambda^2}}.
\end{align}
Furthermore, from \eqref{eq76} and $b>0$ we have $\gamma + \lambda>0$ and~\eqref{eq78}.

Since this equation has a real solution $b$, $\gamma$ must satisfy
\begin{align}
\lambda+ \gamma \geq  M \frac{(P/M)\lambda^2}{1+P\lambda} =\frac{P\lambda^2}{1+P\lambda},
\end{align}
which means
\begin{align}
\gamma \geq   -\frac{\lambda}{1+P\lambda}.
\end{align}
On the other hand, we have $\gamma <0$ from \eqref{eq79}. Hence \eqref{eq77} holds.
\QED

\vspace{0.2cm}

Define a normalized covariance matrix by
\begin{align}
{\bf R}_n&=\frac{1}{(P/M)}\bbE\left[{\bf S}_n{\bf S}_n^T\right] \nonumber  \\
 &=\frac{1}{(P/M)}\left[ \begin{array}{cccc}\bbE[S_n^{(1)} S_n^{(1)}] &\cdots&\bbE[S_n^{(1)}S_n^{(M)}]\\ \bbE[S_n^{(2)}S_n^{(1)}] &\cdots&\bbE[S_n^{(2)}S_n^{(M)}]\\ \vdots&\ddots&\vdots\\ \bbE[S_n^{(M)}S_n^{(1)}] &\cdots& \bbE[S_n^{(M)}S_n^{(M)}]\end{array}\right] \nonumber \\
\label{eq88}
&=\left[ \begin{array}{cccc}r^{(1,1)}_n&\cdots&r^{(1,M)}_n\\r^{(2,1)}_n&\cdots&r^{(2,M)}_n\\ \vdots&\ddots&\vdots\\r^{(M,1)}_n &\cdots&r^{(M,M)}_n\end{array}\right],
\end{align}
where  
\begin{align}
r^{(m,k)}_n\equiv\frac{\bbE[S_n^{(m)}S_n^{(k)}]}{(P/M)}.
\end{align}
For $1\leq m \leq M$, let $H_m$ be the $m$-th column vector of Hadamard matrix ${\bf H}$, and set vector ${\boldsymbol \alpha}_n\equiv[\begin{array}{cccc}\alpha_n^{(1)},\alpha_n^{(2)},\cdots, \alpha_n^{(M)}\end{array}]^T=H_{(n-1\bmod M)+1}$. In addition, we also set $b_n^{(m)}= b_n \hspace{1mm} \alpha_n^{(m)}$ for each $m$ where $\{b_n\}$ is a real sequence. We define a related matrix ${\bf G}_n$ by
\begin{equation}
\label{eq90}
{\bf G}_n = {\bf R}_n -\gamma_n {\bf I}_M,
\end{equation} where $\{\gamma_n\}$ is another real sequence. 

Let $\lambda^{(1)}, \lambda^{(2)},...,\lambda^{(M)}$ be the set of the positive numbers defined in Lemma \ref{bc:lemma3}. We first show by induction that if ${\bf G}_M$ is symmetric positive definite and all column vectors of $M \times M$ Hadamard matrix are eigenvectors of ${\bf G}_M$, then by suitably choosing sequences $b_n, \gamma_n, \beta_n$ for all $n \geq M$, 
matrices ${\bf G}_n$ also satisfy the same properties. In addition, in this case, if  $\lambda_M^{(1)}= \lambda^{(1)}, \lambda_M^{(2)}=\lambda^{(2)},...,\lambda_M^{(M)}=\lambda^{(M)}$, we also have $\lambda_n^{(1)}= \lambda^{(1)}, \lambda_n^{(2)}=\lambda^{(2)},...,\lambda_n^{(M)}=\lambda^{(M)}$ for all $n \geq M$. Here, $\lambda_n^{(m)}$ is the eigenvalue determined by the $[(n+m-2)\bmod M +1]$-th column vector of $M\times M$ Hadamard matrix for each $m=1,2,...,M$. For notation simplicity, denote by $\lambda_n =\lambda_n^{(1)}$, and $\lambda=\lambda^{(1)}$, hereafter.  

We first show that if ${\bf G}_n$ is symmetric definite and $\lambda_n^{(m)}=\lambda^{(m)}$ for $1\leq m\leq M$, then ${\bf G}_{n+1}$ and $\lambda_{n+1}^{(m)}$ satisfying the same property. Denote
\begin{align}
{\bf G}_n\equiv \left[ \begin{array}{cccc}\rho^{(1,1)}_n&\cdots&\rho^{(1,M)}_n\\\rho^{(2,1)}_n&\cdots&\rho^{(2,M)}_n\\ \vdots&\ddots&\vdots\\\rho^{(M,1)}_n &\cdots&\rho^{(M,M)}_n\end{array}\right].
\end{align}
Then from \eqref{eq90}, we obtain
\begin{align}
\label{eq92}
\rho_n^{(m,k)}= r_n^{(m,k)}- \gamma_n \delta(m-k),
\end{align}
where $\delta(n)= 1$ if $n=0$ and $\delta(n)=0$ if $n \neq 0$. Since in our encoding scheme, $X_n$ is given by \eqref{eq10}, and ${\bf R}_n$ and ${\bf G}_n$ are defined by \eqref{eq88} and \eqref{eq90}, respectively, the expected input power at time $n$, $\bbE[X_n^2]$, can be represented by
\begin{align}
\bbE[X_n^2]&=\beta_n^2 \frac{P}{M}{\boldsymbol \alpha}_n^T {\bf R}_n {\boldsymbol \alpha}_n \nonumber \\
&=\beta_n^2\frac{P}{M} \left[{\boldsymbol \alpha}_n^T {\bf G}_n {\boldsymbol \alpha}_n + \gamma_n {\boldsymbol \alpha}_n^T {\bf I}_n {\boldsymbol \alpha}_n\right]  \nonumber \\
&= \beta_n^2 \frac{P}{M}\left[M\lambda + \gamma_n M\right] \nonumber \\
\label{eq93}
&= P \beta_n^2 (\lambda +\gamma_n),
\end{align} where the third equality holds from the fact that ${\bf G}_n {\boldsymbol \alpha}_n = \lambda {\boldsymbol \alpha}_n$ and ${\boldsymbol \alpha}_n^T{\boldsymbol \alpha}_n=||{\boldsymbol \alpha}_n||_2^2=M$.

On the other hand, since the relation between $S_n^{(m)}$ and $Y_n^{(m)}$ is given by \eqref{eq12} with $Z_n=0$ and $\bbE [Z_n^{(m)}]=0$, we obtain that
\begin{align}
\bbE[S_n^{(m)} Y_n^{(k)}] &= \bbE\left[S_n^{(m)} \left(\beta_n \sum_{t=1}^M \alpha_n^{(t)} S_n^{(t)} + Z_n^{(m)}\right)\right]  \nonumber \\
&=\frac{P}{M} \beta_n \sum_{t=1}^M \alpha_n^{(t)} r_n^{(m,t)} \nonumber  \\
&=\frac{P}{M} \beta_n  \sum_{t=1}^M \alpha_n^{(t)} [\rho_n^{(m,t)} + \gamma_n \delta(m-t)]  \nonumber \\
\label{eq94}
&=\frac{P}{M} \beta_n {\boldsymbol  \alpha}_n^T {\boldsymbol \rho}_n^{(m)} + \frac{P}{M} \beta_n \gamma_n  \alpha_n^{(m)},
\end{align}
where the third equality holds from \eqref{eq92}, and
\begin{align}
{\boldsymbol \rho}_n^{(m)}\equiv\left[\begin{array}{cccc}\rho_n^{(m,1)},\rho_n^{(m,2)},\cdots, \rho_n^{(m,n)}\end{array}\right]^T.
\end{align}
From the assumption that ${\bf G}_n$ is symmetric and $\lambda_n$ is the eigenvalue associated with the eigenvector ${\boldsymbol \alpha}_n$ of this matrix, we have
\begin{align}
{\boldsymbol  \alpha}_n^T {\bf G}_n ={\boldsymbol  \alpha}_n^T {\bf G}_n^T= {\boldsymbol \alpha}_n^T \lambda, 
\end{align}
which means 
\begin{align}
\label{eq97}
{\boldsymbol  \alpha}_n^T {\boldsymbol \rho}_n^{(m)} =\lambda \alpha_n^{(m)}.
\end{align}
Substituting~\eqref{eq97} into~\eqref{eq94}, we obtain 
\begin{equation}
\label{eq98}
\bbE[S_n^{(m)} Y_n^{(k)}] =\frac{P}{M}\beta_n (\lambda +\gamma_n) \alpha_n^{(m)}.
\end{equation}
Furthermore, we also obtain
\begin{align}
\bbE[Y_n^{(m)} Y_n^{(k)}]&= \bbE[(X_n + Z_n^{(m)})(X_n+ Z_n^{(k)})] \nonumber \\
&=\bbE[X_n^2] + \bbE[Z_n^{(m)} Z_n^{(k)}] \nonumber \\
\label{eq99}
&= P \beta_n^2 (\lambda +\gamma_n)  + \delta(m-k).
\end{align}
Note from~\eqref{eq9} and $b_n^{(m)}=b_n \alpha_n^{(m)}$ that if
we set $a_n^{(m)}=a_n$ for all $m$, 
our transmission scheme satisfies
\begin{align}
S_{n+1}^{(m)}= \frac{1}{a_n} \left(S_n^{(m)} - b_n \alpha_n^{(m)} Y_n^{(m)}\right).
\end{align}
Therefore, we have
\begin{align}
&\bbE[S_{n+1}^{(m)}S_{n+1}^{(k)}]= \frac{1}{a_n^2} \left(\bbE[S_n^{(m)} S_n^{(k)}] -b_n \alpha_n^{(m)}\bbE[S_n^{(k)} Y_n^{(m)}] \right. \nonumber \\
&\quad \left. -b_n\alpha_n^{(k)} \bbE[S_n^{(m)} Y_n^{(k)}] + b_n^2 \alpha_n^{(m)} \alpha_n^{(k)} \bbE[Y_n^{(m)}Y_n^{(k)}]\right).
\end{align}
Then, 
\begin{align}
\frac{P}{M}r_{n+1}^{(m,k)} &=\frac{1}{a_n^2}\left(\frac{P}{M} r_n^{(m,k)} - b_n \alpha_n^{(m)} \frac{P}{M}\beta_n (\lambda +\gamma_n) \alpha_n^{(k)}   \right. \nonumber \\
&\quad \left. -  b_n \alpha_n^{(k)} \frac{P}{M}\beta_n (\lambda +\gamma_n) \alpha_n^{(m)}\right. \nonumber \\
&\quad \left. +  b_n^2 \alpha_n^{(m)} \alpha_n^{(k)} [P \beta_n^2 (\lambda +\gamma_n)+ \delta(m-k)] \right)  \nonumber \\
&=\frac{1}{a_n^2} \left( \frac{P}{M} r_n^{(m,k)} - 2b_n \beta_n \frac{P}{M} (\lambda+ \gamma_n) \alpha_n^{(m)}\alpha_n^{(k)} \right.  \nonumber \\
\label{eq102}
&\quad \left. + b_n^2 \alpha_n^{(m)} \alpha_n^{(k)} [P \beta_n^2 (\lambda +\gamma_n)+ \delta(m-k)]  \right).
\end{align}
Hence, it holds from~\eqref{eq92} and~\eqref{eq102} that
\begin{align}
&\rho_{n+1}^{(m,k)} + \gamma_{n+1} \delta(m-k) =\frac{1}{ a_n^2} \left(  \rho_n^{(m,k)} + \gamma_n \delta(m-k) \right. \nonumber \\
&\quad - 2 b_n \beta_n (\lambda + \gamma_n) \alpha_n^{(m)} \alpha_n^{(k)} + M b_n^2 \beta_n^2 (\lambda + \gamma_n) \alpha_n^{(m)} \alpha_n^{(k)} \nonumber \\
\label{eq103}
&\quad \left. + \frac{M}{P} b_n^2 \alpha_n^{(m)} \alpha_n^{(k)} \delta(m-k)\right).
\end{align}
Now, if we use
\begin{align}
\label{eq104}
\gamma_{n+1} = \frac{1}{a_n^2} \left(\gamma_n + \frac{M}{P} b_n^2 \right),
\end{align} 
then for all $m, k$ we have
\begin{align}
\label{eq105}
\gamma_{n+1}& \delta(m-k) \nonumber\\
&= \frac{1}{a_n^2} \left( \gamma_n \delta(m-k) + \frac{M}{P}  b_n^2 \alpha_n^{(m)} \alpha_n^{(k)} \delta(m-k) \right).
\end{align}
Combining~\eqref{eq105} with \eqref{eq103}, we obtain
\small
\begin{align}
&\rho_{n+1}^{(m,k)} \nonumber\\
&= \frac{1}{a_n^2}\left[ \rho_n^{(m,k)} -2 b_n \beta_n (\lambda + \gamma_n ) \alpha_n^{(m)} \alpha_n^{(k)} \right. \nonumber \\
& \hspace{2.5cm}\left. \quad + M b_n^2 \beta_n^2(\lambda+\gamma_n) \alpha_n^{(m)} \alpha_n^{(k)} \right] \nonumber \\
\label{eq106}
&= \frac{1}{a_n^2}\left(\rho_n^{(m,k)}  - \left[2 b_n \beta_n(\lambda+\gamma_n)\right.\right.\nonumber\\
& \hspace{2.7cm}\left.\left. - M b_n^2\beta_n^2(\lambda+\gamma_n)\right] \alpha_n^{(m)} \alpha_n^{(k)}  \right).
\end{align}
\normalsize

Now, for all $n \geq M$, we set 
\begin{align}
\label{eq107}
a_n = a= \sqrt{\frac{1+(P/M) \lambda(M-\lambda)}{1+P\lambda}}, 
\end{align}
$b_n=b$ and $\gamma_n=\gamma<0$ where $(b,\gamma)$ is given in Lemma \ref{bc:lemma4}. In order to satisfy the input power constraint, we set $\beta_n$ as follows. 
\begin{align}
\beta_n=\sqrt{\frac{1}{\lambda +\gamma_n}} =\sqrt{\frac{1}{\lambda +\gamma}}. 
\label{mac:eq-new108}
\end{align}
Then, it holds from \eqref{eq93} and \eqref{mac:eq-new108} that $E[X_n^2] = P$ for all $n\geq M$. In addition, we also see from~\eqref{eq78} and \eqref{mac:eq-new108} that
\begin{align}
2 b_n \beta_n (\lambda +\gamma_n)&- M b_n^2\beta_n^2(\lambda+\gamma_n) \nonumber\\
&=2b\sqrt{\lambda+\gamma} - Mb^2\nonumber\\
&= \frac{(P/M) \lambda^2}{1+ P\lambda}.\label{eq109}
 \end{align} 
Substituting~\eqref{eq109} into~\eqref{eq106} we obtain the following recursion:
\begin{align}
\rho_{n+1}^{(m,k)} &=\frac{1+P \lambda} {1 + (P/M) \lambda(M-\lambda)} \rho_n^{(m,k)} \nonumber \\
\label{eq110}
 &\quad - \frac{(P/M) \lambda^2}{1 + (P/M) \lambda(M-\lambda)} \alpha_n^{(m)} \alpha_n^{(k)},
\end{align}
which means
\begin{align}
{\bf G}_{n+1} &=\frac{1+P \lambda} {1 + (P/M) \lambda(M-\lambda)} {\bf G}_n \nonumber \\
\label{eq111}
&\quad - \frac{(P/M) \lambda^2}{1 + (P/M) \lambda(M-\lambda)} {\boldsymbol \alpha}_n {\boldsymbol \alpha}_n^T.
\end{align}
We easily note from~\eqref{eq111} that when ${\bf G}_n$ is symmetric, ${\bf G}_{n+1}$ is also symmetric. Denote $ {\bf H}_n=[{\boldsymbol \alpha}_n\quad {\boldsymbol \alpha}_{n+1}\quad \cdots \quad {\boldsymbol \alpha}_{n+M-1}] $. By our induction assumption, the column vectors of ${\bf H}_n$ are $M$ linearly independent eigenvectors of ${\bf G}_n$.  Furthermore, it holds from~\eqref{eq111} that
\begin{align}
&{\bf H}_{n+1}^T{\bf G}_{n+1}{\bf H}_{n+1}\nonumber\\
&=\frac{1+P\lambda}{1+(P/M)\lambda(M-\lambda)}{\bf H}_{n+1}^T{\bf G}_n{\bf H}_{n+1} \nonumber \\
\label{eq112}
&\hspace{0.6cm} -\frac{(P/M)\lambda^2}{1+(P/M) \lambda(M-\lambda)}{\bf H}_{n+1}^T{\boldsymbol \alpha}_n {\boldsymbol \alpha}_n^T{\bf H}_{n+1}.
\end{align}
Note that since all column vectors of ${\bf H}_n$ are eigenvectors of ${\bf G}_n$, all the column vectors of the matrix ${\bf H}_{n+1}$ are also eigenvectors of ${\bf G}_n$. Hence we has the following eigenvalue decomposition
\begin{align}
{\bf \Lambda}_n={\bf H}^T_{n+1} {\bf G}_n {\bf H}_{n+1},
\end{align} where ${\bf \Lambda}_n =M \mbox{diag}(\lambda^{(2)}, \lambda^{(3)},...,\lambda^{(M)}, \lambda^{(1)})$, which is a diagonal matrix. We also note that
\begin{align}
{\bf H}_{n+1}^T{\boldsymbol \alpha}_n {\boldsymbol \alpha}_n^T{\bf H}_{n+1}&=[{\boldsymbol \alpha}_n^T{\bf H}_{n+1}]^T {\boldsymbol \alpha}_n^T{\bf H}_{n+1}\nonumber\\
&=M^2 \mbox{diag}(0,0,...,0,1)
\end{align}
because ${\boldsymbol \alpha}_{n+M}={\boldsymbol \alpha}_n$, and hence
\begin{align}
{\boldsymbol \alpha}_n^T{\bf H}_{n+1}&= {\boldsymbol \alpha}_n^T\left[\begin{array}{cccc}{\boldsymbol \alpha}_{n+1}& {\boldsymbol \alpha}_{n+2}&\cdots&{\boldsymbol \alpha}_{n+M}\end{array}\right] \nonumber \\
\label{eq116}
&=\left[\begin{array}{cccc}0&0&\cdots&M\end{array}\right].
\end{align}

From \eqref{eq112}--\eqref{eq116}, ${\bf H}_{n+1}^T{\bf G}_{n+1}{\bf H}_{n+1}$ must be a diagonal matrix. Hence, all column vectors of ${\bf H}_{n+1}$are eigenvectors of ${\bf G}_{n+1}$. Moreover, we obtain from \eqref{eq112} that for $1\leq m\leq M-1$,
\begin{align}
\lambda_{n+1}^{(m)}&=\frac{1+P\lambda}{1+(P/M)\lambda(M-\lambda)}\lambda^{(m+1)} \nonumber \\
&\stackrel{(a)}{=}\lambda^{(m)}
\label{eq117}
\end{align}
and
\begin{align}
\lambda_{n+1}^{(M)}&=\frac{1+P\lambda}{1+(P/M)\lambda(M-\lambda)}\lambda-\frac{(P/M)\lambda^2}{1+(P/M)\lambda(M-\lambda)}M \nonumber \\
&=\frac{\lambda}{1+(P/M)\lambda(M-\lambda)} \nonumber \\
&\stackrel{(a)}{=}\frac{1}{1+(P/M)\lambda(M-\lambda)}\left[\frac{1+P\lambda}{1+(P/M)\lambda(M-\lambda)}\right]^{M-1} \nonumber \\
&\hspace{45mm} \cdot \lambda^{(M)} \nonumber \\
&\stackrel{(b)}{=}\lambda^{(M)}
\label{eq118}
\end{align}
where (a) and (b) holds from \eqref{eq87} and \eqref{eq68}, respectively.

Therefore, from~\eqref{eq111},~\eqref{eq117} and~\eqref{eq118}, ${\bf G}_{n+1}$ is symmetric positive definite and $\lambda_{n+1}^{(m)}=\lambda^{(m)}$ if ${\bf G}_n$ is symmetric positive definite and $\lambda_n^{(m)}=\lambda^{(m)}$. 

Next, we show that ${\bf G}_M$ can be derived from 
\begin{equation}
\label{eq119}
{\bf G_1} = \lambda_0 {\bf I}_M
\end{equation} by choosing of parameters $\gamma_n, b_n, \beta_n, a_n, b_n, \lambda_0$ appropriately for $1\leq n \leq M$. In the same way as the case of $n\geq M$, we set $a_n=a, b_n=b, \gamma_n=\gamma$ where $a$ and $(b,\gamma)$ are given by~\eqref{eq107} and Lemma~\ref{bc:lemma4}, respectively. But, we allow that $\lambda_n^{(m)}$ depends on $n$ for $1\leq n\leq M$. Then, in the same way as~\eqref{eq106}, we obtain the following relation:
\begin{align}
&\rho_{n+1}^{(m,k)}\nonumber\\
&=\frac{1}{a^2}\left[\rho_n^{(m,k)}-\big(2b\beta_n(\lambda_n+\gamma) \right. \nonumber\\
&\hspace{2.5cm}\left.\left. -Mb^2\beta_n^2(\lambda_n+\gamma)\right)\alpha_n^{(m)}\alpha_n^{(k)}\right].\label{eq120}
\end{align}
Now, we consider $d_n, 1\leq n\leq M-1$, that satisfies 
\begin{align}
\label{eq121}
2 b \beta_n (\lambda_n +\gamma)- M b^2\beta_n^2(\lambda_n+\gamma) =\left(\frac{1- d_n}{M}\right)\lambda_n.
\end{align} 
Then,~\eqref{eq120} becomes
\begin{align}
\rho_{n+1}^{(m,k)} =\frac{1}{a^2} \left[\rho_n^{(m,k)} - \left(\frac{1-d_n}{M}\right)\lambda_n\alpha_n^{(m)} \alpha_n^{(k)} \right],
\end{align}
which means
\begin{align}
\label{eq123}
{\bf G}_{n+1} = \frac{1}{a^2} \left[{\bf G}_n - \left(\frac{1-d_n}{M}\right)\lambda_n {\boldsymbol \alpha}_n {\boldsymbol \alpha_n}^T\right].
\end{align}
Furthermore, in the same way as \eqref{eq117} and~\eqref{eq118}, we obtain
\begin{align}
\label{eq124}
\lambda_{n+1}^{(m)}=\begin{cases}(1/a^2) \lambda_n^{(m+1)},\hspace{8mm} m=1,2,...,M-1,\\ (d_n/a^2) \lambda_n^{(1)}, \hspace{10mm} m=M.
\end{cases}
\end{align}
From \eqref{eq123} and \eqref{eq124}, we note that ${\bf G}_n$ is symmetric positive definite for $1\leq n\leq M$ if $d_n$ is positive.

We now derive $d_n$ and $\lambda_0$ such that ${\bf G}_M$ has eigenvalues $\lambda^{(1)},\lambda^{(2)},...,\lambda^{(M)}$, which are defined in Lemma~\ref{bc:lemma3}. Note that $\lambda_1^{(m)}=\lambda_0$ for $1\leq m \leq M$. Hence, applying \eqref{eq124} $M-1$ times, we obtain
\begin{align}
 \lambda_M^{(m)} &= \frac{d_{m-1}}{a^{2(M-1)}} \lambda_0 \nonumber \\
 &=\left[\frac{1+P\lambda}{1+(P/M)\lambda(M-\lambda)}\right]^{M-1} d_{m-1} \lambda_0  \nonumber \\
 &=\frac{[1+P\lambda]^{M-1}}{[1+(P/M)\lambda(M-\lambda)]^M} \nonumber \\
\label{eq125}
 &\hspace{1.5cm} \times [1+ (P/M) \lambda (M-\lambda)] d_{m-1} \lambda_0
\end{align}
where $d_0\equiv1$. Since $\lambda$ is the solution of~\eqref{eq87}, \eqref{eq125} means
\begin{align}
\lambda_M^{(m)} =  [1+ (P/M) \lambda (M-\lambda)] d_{m-1} \lambda_0.
\end{align}
Hence, in order to satisfy $\lambda_M^{(m)}=\lambda^{(m)}$, $\lambda_0$ and $d_{m-1}$ must satisfy 
\begin{align}
\label{eq127}
  [1+ (P/M) \lambda (M-\lambda)] d_{m-1} \lambda_0= \lambda^{(m)}.
\end{align}
Since $\lambda_M^{(1)} = \lambda^{(1)} =\lambda$ and $d_0=1$, we obtain
\begin{align}
\label{eq128}
 \lambda_0 =\frac{\lambda}{1+(P/M)\lambda(M-\lambda)}
\end{align}
and 
\begin{align}
\label{eq129}
d_{m-1} = \frac{\lambda^{(m)}}{\lambda}.
\end{align}
On the other hand, it holds from~\eqref{eq68} and~\eqref{eq107} that 
\begin{align}
\label{eq130}
\lambda^{(m)} = a^{2(m-1)} \lambda.
\end{align}
Comparing~\eqref{eq129} with~\eqref{eq130}, we have $d_{m-1} = a^{2(m-1)}$ for all $1\leq m \leq M$. This means that 
\begin{align}
\label{eq131}
d_n = a^{2n},\hspace{1mm} \mbox{for}\hspace{1mm} 1\leq n \leq M-1. \end{align}

To complete the proof, we need to show that~\eqref{eq121} has a positive solution $\beta_n$ for $d_n=a^{2n}$. Note that~\eqref{eq121} has a real solution $\beta_n b$ if 
\begin{align}
\label{eq132}
(\lambda_n + \gamma)^2 \geq (\lambda_n + \gamma) M \frac{1-d_n}{M}\lambda_n, 
\end{align}
i.e.,
\begin{align}
\label{eq133}
(\lambda_n +\gamma) (\gamma + d_n \lambda_n) \geq 0.
\end{align}
From \eqref{eq124}, \eqref{eq128}, and $\lambda_1^{(m)}=\lambda_0$ for $1\leq m\leq M-1$, $\lambda_n$ satisfies
\begin{align}
\lambda_n=\lambda_n^{(1)}&=\frac{1}{a^{2(n-1)}}\lambda_1^{(n)} \nonumber \\
&=\frac{1}{a^{2(n-1)}} \lambda_0 \nonumber \\
&=\frac{1}{a^{2(n-1)}}\left[\frac{\lambda}{1+(P/M)\lambda(M-\lambda)}\right].
\label{eq134}
\end{align}
Therefore, from \eqref{eq107}, \eqref{eq131},  and\eqref{eq134}, we obtain
\begin{align}
\gamma + d_n \lambda_n &= \gamma + a^{2n} \frac{1}{a^{2(n-1)}} \left[\frac{\lambda}{1+(P/M)\lambda(M-\lambda)}\right]  \nonumber \\
&=\gamma + a^2  \left[\frac{\lambda}{1+(P/M)\lambda(M-\lambda)}\right] \nonumber \\
&=\gamma + \frac{1+(P/M)\lambda(M-\lambda)}{1+P\lambda}\nonumber\\
& \hspace{2.5cm} \times
\left[\frac{\lambda}{1+(P/M)\lambda(M-\lambda)}\right] \nonumber \\
\label{eq135}
&=\gamma + \frac{\lambda}{1+P\lambda} \geq  0,
\end{align}
where the last inequality follows from \eqref{eq77}.  On the other hand, since
it holds from \eqref{eq107} that $a^2<1$, 
we have $d_n = a^{2n} <1$. Therefore, it holds from \eqref{eq135} that $\gamma +\lambda_n > \gamma + d_n \lambda_n \geq 0$, which means that 
~\eqref{eq133} also holds. Hence,~\eqref{eq121} has two positive solutions $b\beta_n$ by Vieta's theorem, but we choose smaller $b \beta_n$
to reduce the transmission power. 

Finally, we check that ${\bf R}_1$ is realizable. From~\eqref{eq90} and ${\bf G}_1= \lambda_0 {\bf I}_M$, we have  
\begin{align}
{\bf R}_1&= {\bf G}_1 +\gamma {\bf I}_M =( \lambda_0+ \gamma) {\bf I}_M \nonumber \\
&= \left[\frac{\lambda}{1+(P/M)\lambda (M-\lambda)}+ \gamma\right] {\bf I}_M.
\end{align}
Since it holds for any positive $\lambda$ that $\frac{\lambda}{1+P\lambda}<\frac{\lambda}{1+(P/M)\lambda(M-\lambda)}$, we have from~\eqref{eq135} that
\begin{align}
\frac{\lambda}{1+(P/M)\lambda (M-\lambda)}+\gamma >0.
\end{align}
This means that the initialized random variable $S_1^{(m)}=F_S^{-1}(\Theta_m)$ used in the encoding scheme given in Section~\ref{section3} must satisfy
\begin{align}
S\sim \mathcal{N}\left(0, (P/M) \left[\frac{\lambda}{1+(P/M)\lambda (M-\lambda)}+ \gamma\right]\right).
\end{align}

Now, we evaluate the achievable rates and error probabilities. Our encoding scheme satisfies
\begin{align}
\bbE[X_n^2] = P \hspace{2mm} \hspace{1mm}\mbox{for}\hspace{1mm} n\geq M,
\end{align}
and hence by the Ces\`aro Mean,
\begin{align}
\limsup_{n\rightarrow \infty}\frac{1}{n}\sum_{k=1}^n \bbE[X_k^2] = P. 
\end{align}
This means that the input power constraint is satisfied. Furthermore, for all $n\geq M$, we also have
\begin{align}
\sum_{m=1}^M W_n^{(m)} &\equiv \sum_{m=1}^M \bbE[S_n^{(m)}]^2 \nonumber \\
&= \frac{P}{M} \mbox{tr}({\bf R}_n) \nonumber \\
&= \frac{P}{M}\left( \mbox{tr}({\bf G}_n) + M\gamma \right)\nonumber \\
&=  \frac{P}{M}\left(\sum_{m=1}^M \lambda_m + M \gamma\right) \nonumber  \\
&< \infty.
\end{align}
Hence, we have $W_n\equiv\sup_m W_n^{(m)} < \infty$ since $M$ is finite. Furthermore, since $1 \leq \lambda  \leq M$, we have
\begin{align}
0 < \limsup_{n\rightarrow \infty} a_n =a= \frac{1+(P/M)\lambda(M-\lambda)}{P\lambda+1}<1.
\end{align}
Therefore, since the two conditions in Theorem \ref{bc:thm1} are satisfied, any rate less than the following $R_m^{*}$ is achievable.
\begin{align}
R_m^*&=-\limsup_{n\to \infty}\log a_n^{(m)} \nonumber  \\
&= -\log a \nonumber \\
\label{eq144}
&= \frac{1}{2} \log\left(\frac{1+P\lambda} {1+(P/M)\lambda(M-\lambda)}\right)\equiv R^*.
\end{align}
Hence, the following sum-rate is achievable:
\begin{align}
\sum_{m=1}^M R_m^*&=\frac{M}{2}\log\left(\frac{1+P\lambda}{1+(P/M)\lambda(M-\lambda)}\right) \nonumber \\
&=\frac{1}{2}\log\left(\frac{1+P\lambda}{1+(P/M)\lambda(M-\lambda)}\right)^M \nonumber \\
&=\frac{1}{2}\log\left(1+P\lambda\right),
\end{align}
where $\lambda$ is the biggest solution in $[1,M]$ of \eqref{eq87}.

\section{Relation between AWGN-BCs and AWGN-MACs}\label{section8}
The time-varying coding approach can be applied to the AWGN-MAC (multiple access channel) with feedback. It is shown in \cite{key10} that the time-varying coding scheme can achieve the linear-feedback sum-capacity for AWGN-MACs \cite{key15} as with the Kramer code \cite{key2} and the LQG code \cite{key4}. 
Let $R_{{\rm MAC}}(M,P)$ denote the achievable symmetric sum-rate by the time-varying code \cite{key10} for $M$-sender AWGN MACs with feedback where each encoder has power constraint $P$. Then, it is shown in \cite[Theorem III]{key10} that  
\begin{align}
\label{eq149}
R_{{\rm MAC}}(M,P)=\frac{1}{2}\log(1+MP\lambda),
\end{align}
where $\lambda$ is the biggest solution of
\begin{align}
(1+MPx)^{M-1}=(1+Px(M-x))^M.
\end{align}
Comparing~\eqref{eq149} with Theorem \ref{bc:thm2}, we note that
\begin{align}
R_{{\rm BC}}(M,P)=R_{{\rm MAC}}(M, P/M).
\end{align}

This shows that when we use the time-varying code under the same sum-power constraint $P$, the achievable sum-rate for MAC is equal to the one for BC. This relation between MAC and BC is already pointed out in \cite{key4} and \cite{key25}. 
From our results, we note that the posterior matching scheme can also attain this duality between MAC and BC.

\section{Conclusion}
We proposed a general coding scheme based on the posterior matching  for AWGN-BCs with feedback, and we derived the achievable rate region and the decoding error probability of the proposed scheme.
 Then, we showed that a variant of the Ozarow-Leung coding scheme can be obtained as a special case of our scheme. Furthermore, we clarified how to realize the posterior matching for the physically non-degraded symmetric AWGN-BCs, and we showed the proposed coding scheme can attain the linear-feedback sum-capacity for these symmetric AWGN-BCs.

An interesting further research topic is to find a good sequences $a_n^{(m)}, b_n^{(m)}$ to attain good performance for more general settings treated in \cite{key5} and \cite{key6}. 
\section*{Acknowledgment}
The authors thank the associate editor and reviewers for their helpful comments.
This work is supported in part by JSPS Grant-in-Aid for Scientific Research, No.~25289111.



\profile{Lan V.~Truong}{was born in Quang Binh province, Vietnam. He received the B.S.E. degree in Electronics and Telecommunications from Posts and Telecommunications Institute of Technology (PTIT), Hanoi, Vietnam in 2003. After many years of working as an operation and maintenance engineer (O\&M) at MobiFone Telecommunications Corporation, Hanoi, Vietnam, he resumed his graduate studies at School of Electrical and Computer Engineering (ECE), Purdue University, West Lafayette, IN, United States and obtained the M.S.E. degree in 2011. From 2013 to June 2015, he was an academic lecturer at Department of Information Technology Specialization (ITS), FPT University, Hanoi, Vietnam. Presently, he is Ph.D. student at Department of Electrical and Computer Engineering (ECE), National University of Singapore (NUS).  His research interests are information theory and its applications. 
}

\profile{Hirosuke Yamamoto}
{was born in Wakayama,
Japan, in 1952. He received the B.E. degree from Shizuoka University,
Shizuoka, Japan, in 1975 and the M.E. and Ph.D. degrees from the University
of Tokyo, Tokyo, Japan, in 1977 and 1980, respectively, all in electrical
engineering. In 1980, he joined Tokushima University. He was an Associate
Professor at Tokushima University from 1983 to 1987, the University of
Electro-Communications from 1987 to 1993, and the University of Tokyo
from 1993 to 1999. Since 1999, he has been a Professor at the University
of Tokyo and is currently with the Department of Complexity Science and
Engineering at the university. In 1989-1990, he was a Visiting Scholar
at the Information Systems Laboratory, Stanford University, Stanford, CA.
His research interests are in Shannon theory, data compression algorithms,
and information theoretic cryptology.
Dr. Yamamoto served as the Chair of IEEE Information Theory Society
Japan Chapter in 2002-2003, the TPC Co-Chair of the ISITA2004, the TPC
Chair of the ISITA2008, the President of the SITA (Society of Information
Theory and its Applications) in 2008-2009, the President of the ESS
(Engineering Sciences Society) of IEICE in 2012-2013, an Auditor of IEICE in 2016-2017, an Associate Editor for Shannon Theory, the IEEE Transactions on Information Theory in 2007-2010, Editor-in-Chief for the IEICE Transactions on Fundamentals
of Electronics, Communications and Computer Sciences in 2009-2011. He is a Fellow of the IEICE and IEEE.} 

\end{document}